\documentclass[twocolumn,secnumarabic,amssymb,nobibnotes, aps, prd]{revtex4}
\usepackage{graphicx}
\usepackage{multirow}
\usepackage{epstopdf}
\usepackage{amsmath}

\def\lsim{\buildrel{\scriptscriptstyle <}\over{\scriptscriptstyle\sim}}
\def\gsim{\buildrel{\scriptscriptstyle >}\over{\scriptscriptstyle\sim}}
\newcommand{\br}{\begin{eqnarray}}
\newcommand{\er}{\end{eqnarray}}
\def \sq{\tilde q}
\def \gl{\tilde g}
\def\t1{\tilde t_1}
\def \CH{{\tilde\chi}^{\pm}}
\def \N0{\tilde\chi^0}
\def \MET{E{\!\!\!/}_T}
\def \inv{^{-1}}
\def\bul{\bullet}

\begin{document}

\title{Event-shape selection cuts for supersymmetry searches at the LHC 
with 7 TeV energy}%
\author{Monoranjan Guchait and Dipan Sengupta}%
\affiliation{Department of High Energy Physics, Tata Institute of 
Fundamental Research, Mumbai, India} 
\date{March, 2011}%
\begin{abstract}
We investigate the prospects of supersymmetry searches at the LHC with 
7 TeV energy. A new set of selection cuts is proposed
based on event shapes to control backgrounds. 
Our preliminary studies show that it is possible to  
minimize backgrounds to a significantly low level and a conservative 
estimate suggests that the mass reach 
can be extended to $\sim$1.1 TeV for luminosity ${\cal L=}$1fb$^{\inv}$ 
with a reasonable signal-to-background ratio. 
\end{abstract}
\maketitle

\section{Introduction}
For more than a few decades it has been  well known that 
supersymmetry(SUSY) is one of the most promising candidates for  
beyond standard model phsics. 
Since March 2010, LHC has been running with a center-of-mass energy 7 
TeV and is expected to accumulate 1fb$\inv$ data by the 
end of 2011. In both the experiments, ATLAS and CMS are heavily engaged to 
look for a SUSY signal in the collision data.  
From negative searches CMS(ATLAS) have predicted a lower bound on 
gluino mass, $m_{\gl} \gsim$ 650(700)GeV for almost degenerate squark 
mass($m_{\sq}$)~\cite{cms,atlas} in the framework 
of the constrained minimal SUSY model (cMSSM).
It is now of general interest to address the discovery potential of 
SUSY at this low energy(7 TeV) and low luminosity(1fb$\inv$) 
option in the LHC experiment. 
Already quite a few studies have been  
carried out along this direction~\cite{tata,pnath,kane}.
\\
\vspace{.2mm}
In this present study, we revisit the discovery reach for SUSY at 7 TeV 
energy with a special effort to control  
standard model(SM) backgrounds by implementing 
a different search strategy. 
In hadron colliders, sparticle 
production, because of its long cascade decay chain 
results in events with a multiple number of leptons and jets along 
with missing   transverse energy($\MET$) due to the presence of 
neutralino($\N0_1$),
assumed to be the lightest sparticle.
Since they originate from massive $\gl$, $\sq$ states, therefore
the final states consist of harder objects. Hence, SUSY is characterized
by high transverse momentum($p_T$) events with higher multiplicities. 
We introduce event-shape variables~\cite{salam} 
to exploit these special features of SUSY events to discriminate 
signal and backgrounds. Moreover, we also try to construct an additional cut
based on $p_T$ of jets in the final state. Feasibility  
of these new sets of cuts are examined by analyzing the   
following final states, which are thought to be discovery 
channels for SUSY at the LHC:
\\
$\bullet$ a single lepton + jets(1$\ell$),\\
$\bul$ di-leptons+ jets(2$\ell$),\\
$\bul$ jets + $\MET$.
\\
\vspace{0.2mm}
In a hadron collider machine, measurement of $\MET$ is a nontrivial task
due to the presence of other nonphysics sources of $\MET$.
However, current studies show that $\MET$ 
performance in the detector is better than it was thought to be~\cite{cmsmet}.
Nevertheless, we investigate the detection possibility of SUSY signal 
with and without $\MET$~\cite{tata2} in the final state.
\\
As is the practice, we simulate signal event in the framework of a minimal 
supergravity(mSUGRA) based SUSY model described by 
four parameters,
$
m_0, m_{1/2}, A_0, \tan\beta
$
and  sign$(\mu)$ at the GUT scale. Here   
$m_0, m_{1/2}$ are the unified masses of scalars and fermions, respectively,
$A_0$ is the trilinear coupling, $\tan\beta$ is the ratio of two 
vacuum expectation values and  $\mu$ is the Higgs mass
parameter. The mSUGRA model is very severely constrained by many low 
energy experimental data, like direct bounds on sparticles as 
well as from dark matter experiments~\cite{pnath,dighe}. Instead of
testing the mSUGRA model against all those constraints, we simply use
restricted parameter space obtained by 
other authors~\cite{pnath} and select a few sets 
of parameters to a simulate signal.
The sparticle masses at the electroweak scale are obtained by using 
the renormalization group 
evolution performed by SuSpect, and decay branching 
ratios of sparticles are calculated using the SUSY-HIT package~\cite{suspect}.
In Table I, four sets of parameters(P1-P4) are presented for fixed values 
of $A_0$=0, $\tan\beta$=45 and sign($\mu$)=+1. 
Masses of $\gl$ and $\sq$ are shown along 
charginos and neutralinos. In all cases (P1-P4) 
the lighter chargino state($\CH_1$) turns 
out to be the next-to-lightest supersymmetric particle(NLSP).
\begin{table}
\begin{ruledtabular}
\begin{tabular}{lllll}
& P1  & P2 & P3 & P4  \\
\hline 
$m_0$ & 500  & 1500 & 500 & 450\\
\hline
$m_{1/2}$ & 200  & 200 & 400 & 500\\
\hline 
$m_{\gl}$ &524  & 575 & 954 & 1161 \\
\hline
$m_{\sq}$ & 660  & 1535 & 981 & 1133  \\ 
\hline
$m_{\CH_{1,2}}$ & 142,296  & 126,241 & 308,515 & 391,623\\
\hline
$m_{\N0_{1,2}}$ & 78,143  & 76,130 & 164,309 & 207,392\\
\hline
$m_{\N0_{3,4}}$ & 274,295  & 196,240 & 499,514 & 610,623\\
\hline
$\sigma$(pb) &2.5  &0.32 &0.08 & 0.018\\
\end{tabular}
\caption{ Masses(in GeV) of SUSY particles for four sets of 
$m_0, m_{1/2}$ and fixed values of $A_0$=0, $\tan\beta$=45,
sign($\mu$)=+1. The total cross sections($\sigma$) for SUSY particle 
production are in the last row.}
\end{ruledtabular}
\end{table}

\section{Signal and Background}
At the LHC, pairs of colored 
sparticles, viz gluino-gluino, gluino-squark, and 
squark-squark, are produced copiously and 
the corresponding total leading order(LO) cross sections are 
presented in the last row in Table I for each of the parameter space.
We set both the renormalization and factorization scales to $Q^2 = \hat s$,
defined to be the center-of-mass energy in the partonic frame,  
and CTEQ6L~\cite{cteq} is used for parton distribution function.
The dominant SM backgrounds are, 
$\rm t\bar t$+jets, W/Z + jets, $\rm t \bar t W$+jets, tbW+jets, QCD.
In addition, we also check contributions due to WW+jets, ZZ+jets 
and WZ+jets. 
Since, the background to signal cross section ratio 
are of several orders of 
magnitudes, therefore, to identify the SUSY signal, one needs to 
suppress backgrounds a by huge amount, which is not a formidable task 
by a suitable choice of kinematical selection cuts~\cite{ptdr}.
However, a continuous effort is always in process to control 
SM backgrounds to achieve a better 
signal sensitivity~\cite{rogan}. For example, a new   
variable, namely $\alpha_T$, was proposed to suppress backgrounds, 
mainly QCD di-jet events~\cite{lisa}.
In this current study, our goal is to
find a better method to deal with SM backgrounds.
We propose a new strategy to suppress 
backgrounds
by using well-known event-shape variable, namely, transverse thrust(T).
In addition, we define another new variable
taking the ratio($R_T$) of a scalar sum of transeverse momentum of
lowest number of jets over the sum of all jets present in the
event. To the best of our knowledge these two cuts are never used 
in SUSY searches in the hadron colliders. We observed, which are discussed 
below, that these two cuts   
play a very important role in isolating SM backgrounds 
leading discovery reach signal cross section limited. Importantly, these
two cuts are also very easy to implement experimentally once four 
momenta of 
jets are reconstructed.       
\\
The event generator {\tt PYTHIA6}\cite{pythia} is used to generate
signal events and background processes due to $t \bar t$, WW,WZ,ZZ 
and QCD.
The $t \bar t$ and QCD
backgrounds are generated by slicing the entire phase space in
various $\hat p_T$ bins, where $\hat p_T$ stands for the transverse 
momentum of final state partons in the partonic center-of-mass frame.
For QCD, we present results from $\hat p_T \ge$200 GeV onwards
since contribution due to low $\hat p_T$ bins are expected to be
negligible because of strong cuts as can be seen later. 
The hard scattering process consisting of more than two particles in the
final state, like $t \bar t$+jets, W/Z+jets, $t\bar t$W+jets tbW+jets, 
are simulated using {\tt ALPGEN}~\cite{alpgen} based on the  
matrix element(ME) calculation and  
subsequently  passed through {\tt PYTHIA6} for parton showering(PS). 
While generating 
events by  {\tt ALPGEN}, initial selection cuts of $p_{T}\ge20 GeV$ and 
pseudorapidity, $|\eta|\le 3$ are applied.  We adopt MLM
matching\cite{mlm} to avoid double counting while doing parton showering(PS) 
after performing matrix element(ME) calculation. In the MLM 
matching, we use jthe et $p_T$ threshold to
20 GeV and $|\eta|\le$3 while keeping the default value of $\Delta R$=0.7 
for jet and parton separation. Finally, we multiply matching 
efficiency with the accepted efficiency to obtain event rates. The jets 
are reconstructed by
{\tt FastJet}~\cite{fastjet} using the anti-$K_T$~\cite{antikt} cone algorithm
and are preselected with minimum cuts $p_T^j\ge$50~GeV and $|\eta_j|\le$3. 
The total $\MET$ of the event is calculated out 
of the momentum 
of all visible particles present in the event with $p_T\ge$1 GeV and 
$|\eta|\le 3$. 
\begin{figure*}[t]
\centering
\includegraphics[height=3.0in,width=3.5in]{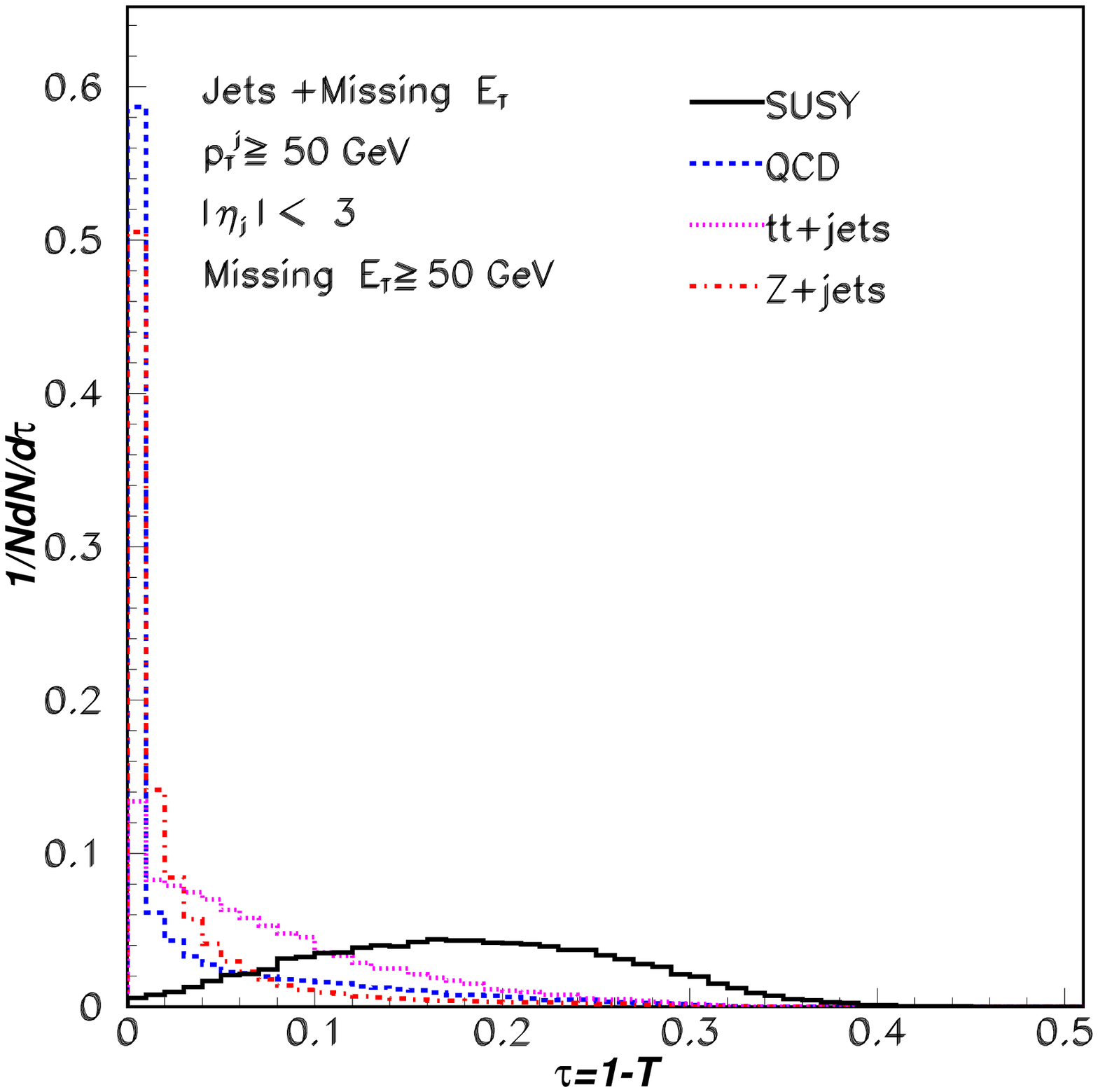}
\includegraphics[height=3.0in,width=3.5in]{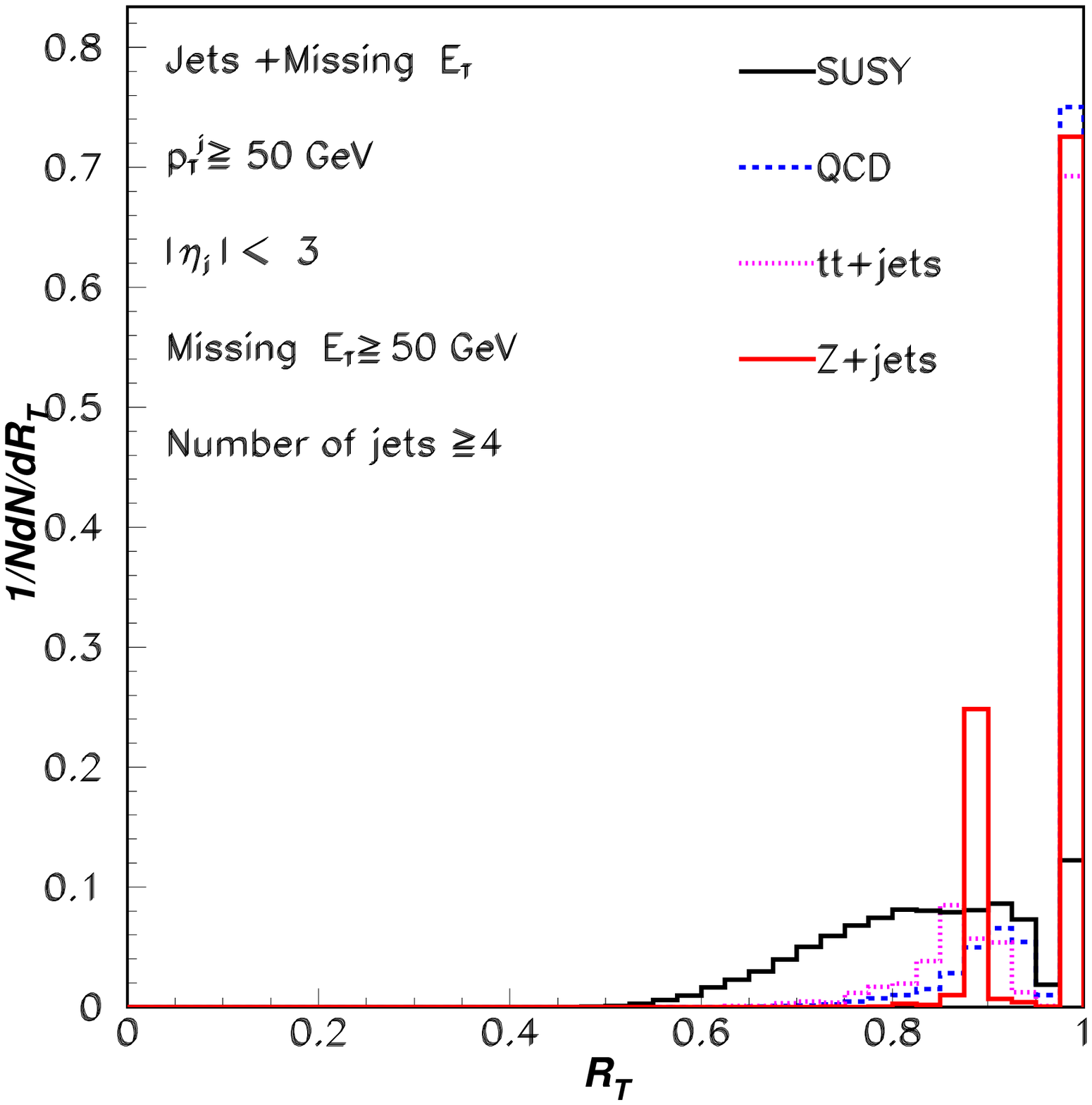}
\caption
{The transverse thrust(left) and $R_T$(right) distributions(normalized to 
unity) for the signal and the backgrounds with selection 
cuts as shown in the figure.}
\label{fig:fig1}
\end{figure*}
\\
In the following we, describe cuts used in the simulation.
{$\bul$}Leptons(C1): Leptons, both electron($e$) and muon($\mu$) are selected 
with $p_T \ge$10~GeV and $|\eta|\le$3. In the case of the single lepton 
final state, we apply $p_{T}\ge 20$GeV.
Isolation of the lepton is ensured by looking at the total transverse
 energy, $E_T^{AC}\le$20\% of the $p_T$ of the lepton, where $E_T^{AC}$ is
 the scalar sum of transverse energies of jets close to the 
lepton satisfying $\Delta R(\ell,j)\le$0.2.
\\
$\bul$Event-Shape variable(C2):
These variables describe the 
shape of the event which is related with the geometry of the final state
and, importantly, are safe to use as observables because of its stability 
against any singularity.
For instance, the CMS has used event-shape variables to test models 
for QCD multijet production by analyzing collision data~\cite{evtshape}.
As discussed before, multiplicity of
the final state, which determines the shape of the event, is 
very different in SUSY and backgrounds.
This encouraging observation leads us to use event-shape variables
to separate out the signal from the debris of SM backgrounds. 
There are many event-shape variables out of which we use  
only the transverse thrust defined in the ($x$-$y$) plane transverse 
to the beam direction, which is along the $z$-axis.
The transverse thrust is defined as~\cite{salam},
\br
{\rm T} = {\rm max}_{n_T}\frac { \sum_i |\vec q_{T,i} . \vec n_T| } 
{\sum_i q_{T,i} },
\label{eq:tht}
\er
where the sum runs over all objects in the event,
$\vec q_{T,i}$ is the transverse component of each object and 
$\vec n_T$ is the transverse vector that maximizes this ratio.
Notice that because of the normalization of $\rm T$ 
by the hard scale of the event, the effect of systematic uncertainties 
related with measurements is expected to be less. 
In Fig. 1, in the left panel, we display the distribution 
for $\tau= 1- {\rm T}$,  
of signal and the backgrounds for jet plus $\MET$ final state. 
This distribution is subject to preselection jet cuts 
i.e,$p_T^j\ge$50~GeV and 
$|\eta|_j\le$3 in conjunction with $\MET\ge$50 GeV.
It clearly demonstrates that events having low 
multiplicity, like high $p_T$ QCD and Z($\to \nu\bar\nu$)+jets processes, 
are distributed near $\tau\sim0$($T \sim 1$) 
whereas $t \bar t$+jets with higher multiplicities are distributed towards
comparatively higher(lower) values of $\tau$(T).
As expected, signal events with more multiplicities 
are distributed towards larger values of $\tau$. 
This novel feature of transverse thrust between 
signal and background events is exploited to suppress the latter. 
For instance, selecting events by demanding $T \le 0.9$($\tau \ge 0.1$), 
QCD background and W/Z+jets are suppressed  by more than 90\% whereas 
$t \bar t$ and the signal are less affected. 
\\
${\bul}$ $R_T$ (C3): 
Events are accepted requiring a number of preselected 
jets, $n_j \ge n_j^{min}$, where $n_j^{min}$ is the lowest number
of jets events should have at the least and is the input for event 
selection. Now   
we construct a ratio($R_T$) between 
the scalar sum of $p_T$ for the required lowest number of jets ($n_{j}^{min}$) 
in the event and the  
scalar sum of $p_T$ of all jets($n_j$) in the same event, 
i.e,
\br
R_{T} = \frac{\sum_1^{n_{j}^{min}} p_T^{j_i}}{H_T}
\label{eq:RT}
\er
with $H_T = \sum_{1}^{n_j} p_{T}^{j_i}$.
It is obvious that the behavior of this variable is strongly connected 
with the multiplicity as well as the hardness of the jets   
in the event. 
The $R_T$ becomes exact unity for those events with $n_j=n_j^{min}$, 
otherwise it is expected to be away from unity for those events with  
$n_j >> n_j^{min}$.
In Fig. 1, we present in the right panel the 
distribution of $R_{T}$ of signal along with backgrounds subject to 
jet selection cuts along with the requirement of $n_j^{min}=$4.
The peak suggests that most of the events in the backgrounds are 
dominantly 4-jet events. 
The dip indicates events with relatively softer jets, 
$p_T^j\sim$50~GeV(recall that preselection cut on jets $p_T\ge$50~GeV) 
along with $n_j$ just above 
$n_j^{min}$, are present predominantly in backgrounds). 
On the other hand, for the case $n_j >> n_j^{min}$ 
and for comparatively harder jets, which is the case for signal events,
$R_T$ is mostly distributed below 0.85.
Evidently, the ${\rm R}_{T}$ distribution provides us another very robust 
tool towards the suppression of SM backgrounds by a substantial amount.
For instance, restricting $R_T \le $0.85 enables to remove backgrounds 
by $\sim$70-80\%  or or more with a 10-20\% loss in signal events.
\\
$\bullet$ $H_T$(C4): 
As mentioned before, the signal events are expected to be harder  
in comparison to background events. Therefore, a cut on $H_T\ge$900~GeV 
is very effective to get rid of a good fraction of background events.
\\
$\bul$ $\MET$ cut(C5): In all backgrounds, the   
lepton mainly comes from $W$ decay except for QCD, hence the transverse mass 
between the lepton and $\MET$ is expected to be bounded by 
the $W$ mass, for the single lepton case, in particular.
Hence, events with a single lepton case is expected to suffer 
due to a cut on transverse mass, 
$M_T = \sqrt{2 E_T^{\ell} \MET (1 - \cos\phi(\ell,\MET}))\ge$60
GeV~\cite{paige}, where $\phi$ is the azimuthal angle between the lepton and 
the $\MET$ direction. Beside this $M_T$ cut, which is applicable only 
for one lepton final state, a cut on $\MET \ge$150~GeV for all cases 
turns out to be a good criterion to reject backgrounds.
\\
{\underline{$\bullet$ {\bf {One lepton + jets(1$\ell$):} }}}
The event with the final state having only a single lepton($e$ or $\mu$) along 
with at least 3 jets($n_{j}^{min}$=3) are simulated
applying a sequential set of cuts C1-C5 and the results are 
presented in Table II using P1 parameter space for signal.
The second and third columns display the raw production cross sections and 
the simulated number of events($N_{{\rm EV}}$), respectively, for each process.
The numbers in the third column corresponding to $t \bar t$+jets and W/Z+jets  
represent the number of events with jet-parton MLM matching.
We observe 
that, in general, about 20\%-25\% of the events 
are lost due to the isolation requirement of the lepton, with the 
exception of the QCD background with more($\sim$50\%) loss.  
In the 5th column, the numbers after thrust(Eq.\ref{eq:tht}) cut(C2) 
$T \le $0.90, clearly indicate that  
the QCD background is eliminated by a huge amount 
whereas other backgrounds suffer
by about 50\%-70\% with a modest loss in the signal event.
Requirement of at least 3 jets with a restriction on 
$R_{T} \le 0.85$(C3) results in a substantial suppression of  
backgrounds, particularly processes with low jet multiplicity.  
Subsequently, a cut(C4) on $H_T\ge$900 GeV removes a significant fraction 
of backgrounds.
The resulting cross sections($\sigma_{0\MET}$) due to cuts C1-C4 i.e 
without the $\MET$ requirement turn out to be 179fb for the signal and 
101fb for the total background leading to S/$\sqrt{B}$ $\sim$ 17. 
However, for other sets of SUSY  
parameters P3-P4, this ratio is $\lsim$ 1. 
We check that the contributions due to backgrounds    
${\rm t\bar t W}$+jets, tbw+jets and VV+jets(V=W,Z) are negligible after 
applying all cuts.
\begin{table}
\begin{ruledtabular}
\begin{tabular}{llllllllll}
\hline
&$\sigma$(pb) & $N_{EV}$ & C1 & C2 & C3 & C4 & C5 
& $\sigma_{\MET}$(fb)    \\
\hline
$\gl\gl$ & 0.49 & 10K & 2425 & 1853 & 1669 & 423 & 142&7\\
\hline 
$\gl\sq$ & 1.37 & 15k & 3468  & 2656 &2423 & 1068&397 & 36 \\
\hline 
$\sq\sq$ & 0.65 & 10k & 2197 & 1648 & 1463 & 933 & 417 
&27.1\\ 
\hline
$t\bar t$(5-200)&89.7 & 100K & 19835 & 8208 &1256 & 13 &2&1.8 \\
$t\bar t$(200-500)& 10.2 & 50K & 12987 & 2678 & 1003 & 35 &2 &0.4\\
$t\bar t$(500-$\inf$) & 0.1 & 20K & 5260 & 329 & 135 & 86 &15 &0.08\\
\hline
$t \bar t$ + 1 jet & 34.8 & 13990 & 3457 &1000 & 300 &1 &0  &0 \\
$t \bar t$ + 2 jet & 13.5 & 4409 & 1252 & 492 & 233 & 3 &0 &0 \\
$t \bar t$ + 3 jet & 3.96 & 2840 & 827 &450 & 292 & 10 & 2 & 0.37\\
\hline
W+1 jet& 12000  & 640283 & 3592 & 1027 & 0 & 0 &0 &0 \\
W+2jet & 3000 & 77189 & 11211 & 3332 &0 & 0 &0 &0\\ 
W+3jet & 660 & 53119 & 17860 & 6054 & 113 & 0 &0 &0  \\
W+4jet & 133 & 17447 & 9198 & 3778 & 678 & 22 &0 &0  \\
\hline
Z+1jet & 3990. & 312652 & 1276 & 340 &1 &0 &0 &0  \\
Z+2jet & 988 & 112032 & 8217 & 2348 & 6 &0 &0 &0  \\
Z+3jet & 213 & 48855 & 8736 & 3076 & 196 & 2 &0 &0  \\
Z+4jet & 66 & 13481 & 3320 & 1407 & 264 & 8 &0 &0    \\
\hline
QCD & & & & & & & & & \\
200-300 & 6868 & 7M & 22284 & 2225 &685 &19 &0&0 \\
300-500 & 837 & 1M & 3626 &328& 157 & 59 & 0 &0 \\
500-800 & 40.3 & 1000K & 438 & 27& 8 &8&0 &0  \\
800-1500 & 1.55 & 50K & 275 & 13 & 3 &3 &0 &0 \\
1500 - $\inf$ & 0.003 & 20K & 121 & 4 & 1 & 1 &0 &0 \\
\end{tabular}
\caption{ Event summary for the signal and the backgrounds after each set  
of cuts(C1-C5) described in the text for the single lepton(1$\ell$) 
case. For QCD and $t \bar t$+jets, events are simulated for 
different $\hat p_T$ bins as shown.}
\end{ruledtabular}
\end{table}
Finally, $\MET$ is considered to suppress the remaining backgrounds to 
gain in signal-to-background ratio. Demanding(C5) the transverse 
mass($M_T$) between the lepton 
and $\MET$, $M_T\ge$60~GeV and $\MET \ge$150 GeV 
it is possible to further isolate backgrounds to an almost negligible 
level( $\sigma_{\MET}\sim$2.65 fb).
Note that due to the cuts C4 and C5, the signal suffers substantially 
because of comparatively lighter $\gl$ and $\sq$ masses. However,
for higher masses, the losses are minimal.
In Table III, we show the total signal(P1-P4) and background 
cross sections after cuts C1-C4($\sigma_{0\MET}$) and C1-C5($\sigma_{\MET}$).
This table predicts that the single lepton channel 
alone can explore $\gl$ and $\sq$ of masses up to $\sim$1 TeV for   
${\cal L}=$1fb$\inv$ requiring a signal(S) to a background(B) ratio,
$S/\sqrt{B} \ge $ 5.
\\
{\bf \underline {$\bullet$ Di-lepton +jets(2$\ell$): }}
The final state with two leptons of any type and of any charge
combinations are accepted along with at least 3 jets(=$n_j^{min}$).  
In the simulation, a softer cut on the lepton, $p_T^\ell\ge$10 GeV
($\ell$=e,$\mu$) is 
imposed along with the isolation requirement.
In this case one of the most dominant backgrounds appears to be due to 
Z+jets. A good fraction of Z+jets events can be removed by  
an additional requirement on the di-lepton invariant mass,   
$m_{\ell^+\ell^-} \ge$10~GeV and $m_{\ell^+\ell^-} \ne$70-120~GeV. 
We find that the patterns of background suppression due to 
cuts C2-C4 are more or less the same as a single lepton case.   
In Table III, we present the   
total cross section for both the signal and the backgrounds with C1-C4 cuts
($\sigma_{0\MET}$) and including $\MET$ cut(C1-C5) 
($\sigma_{\MET}$).
This table depicts that, without $\MET$ cut it is difficult to 
observe any signal event in the di-lepton channel except for 
light $\gl$ and $\sq$ masses(P1). Therefore, including $\MET$ cut(C5) 
it may be possible to detect signal events for $\gl$ and $\sq$ 
masses $\sim$700 GeV with ${\cal L}=1$fb$\inv$.
\\
{\bf\underline {$\bullet$ Jets +$\MET$: }}
It is well known that SUSY searches in jets plus the missing energy 
channel offer the highest reach of gluino and squark masses~\cite{cms,atlas}. 
We revisit this final state to examine the effects of  
transverse thrust(Eq.\ref{eq:tht}) and $R_T$(eq.\ref{eq:RT}) selection
cuts in tandem with the other set of cuts.
A substantial number of background events($\sim$50\%-80\% or more) 
are rejected by thrust cut(C2) costing about 20\% of the signal events. 
Furthermore, events with at least 4(=$n_{j}^{min}$) jets  
with a softer cut on the 4th subleading jet, $p_T^{j_4}\ge$70 GeV are
selected and a requirement on  $R_T\le$0.85  
turns out to be very useful 
to get rid of a good amount of W/Z+jets and $t\bar t W$+jets events.
However, C1-C5 cuts are not enough to
kill backgrounds completely, a good amount of $t\bar t$+jets remains.  
Therefore, in addition, we adopt a selection cut 
on the transverse mass $m_T^{j_1 j_2}$, between two leading jets.    
The demand of $m_T^{j_1j_2}\ge$450~GeV is very effective
in suppressing  $t\bar t$+jets events by almost a factor 10, of course, 
with a substantial loss($\sim$60\%-70\%) of signal cross section 
as well. However, for higher masses of $\gl$ and $\sq$ 
i.e for parameter sets P3 and P4, we find this loss is less than 5\%. 
The cut $H_T\ge$1050~GeV removes the remaining 
background events from W/Z+jets, whereas  
high $p_T$ QCD and $t\bar t$ events are eliminated    
by  the $\MET \ge$150 GeV cut. We find after all cuts the total background 
cross section mainly due to the $t \bar t$ and QCD is 
$\sim$ 3.7fb out of which $t\bar t$+jets contributes about 70\%.    
In Table III the total signal (P1-P4) and background 
cross sections are presented. It reflects that at 
$\gl$ and $\sq$ masses up to $\sim$1.1~TeV can be explored 
with 1fb$\inv$ luminosity. We find that the signal efficiency for 
low mass point (P1) is about 10\% and for higher mass(P4) it turns 
out to be $\sim$30\%-40\%. 
\\
\section{Conclusions}
We investigate the detection possibility of the SUSY 
signal in three different 
channels by implementing a set of cuts few of which
were not used in earlier analysis. Our study shows that one of the 
eventshape variables, namely, transverse thrust, and another observable 
$R_T$ play a 
very significant role in disentangling backgrounds from the signal 
events.  
Interestingly, both these observables, T and $R_T$ are
useful in isolating backgrounds consisting of higher 
multiplicity, unlike the case the 
for $\alpha_T$~\cite{lisa} variable which works for di-jet events only. It 
implies 
that these variables, $T$ and $R_T$ are complimentary to $\alpha_T$.  
The added advantage is that  being dimensionless variables, 
measurements of $T$ and $R_T$ involve less systematic uncertainties.   
Our conservative estimate predicts that in the   
single lepton channel and as well as the jet plus $\MET$ channel, 
it is possible to achieve a reasonable signal-to-background ratio for 
$\gl$ and $\sq$ masses up to $\sim$ 1.1 TeV whereas the di-lepton channel 
alone is not very encouraging.
\begin{table}
\begin{ruledtabular}
\begin{tabular}{|l|llllll}
& Total Bg & P1  & P2 & P3 & P4  \\
\hline
1$\ell$($\sigma_{0\MET}$)        & 101 & 179 & 20 & 7 & 2 \\
1$\ell$($\sigma_{\MET}$) & 2.65 & 70. & 8 & 5 & 1.3 \\
\hline
2$\ell$($\sigma_{0\MET}$) & 5.43 & 56 & 7 & 2 & 0.5 \\
2$\ell$($\sigma_{\MET}$) & 0.97 & 31 & 4 & 1.8 & 0.5 \\ 
\hline
Jets($\sigma_{\MET}$) & 3.7 & 271 & 32.5 & 21.8 & 4.63 \\
\end{tabular}
\caption{Total signal(P1-P4) and background cross sections(fb) after
cuts C1-C4($\sigma_{0\MET}$) and C1-C5($\sigma_{\MET}$) for the single 
lepton(1$\ell$), di-lepton(2$\ell$) and jets plus $\MET$ case.}
 \end{ruledtabular}
\end{table} 
It is to be noted that our conclusion is based on LO signal cross sections  
whereas in background evaluation the  
higher order effects are taken into account to a certain extent by 
considering hard emission of partons(jets), which is the real part of the 
NLO correction. We check using 
{\tt Prospino}~\cite{prospino} that 
NLO cross sections for SUSY are about 50\%-70\% higher than LO cross 
sections resulting in an enhancement in signal-to-background ratio 
than our present conservative estimate. 
Clearly, the discovery reach is signal rate limited 
rather than background limited, which is almost negligible after 
all cuts. Evidently, this is a very powerful
result based on our search strategy, which can be implemented experimentally
very easily. For illustration purposes, we mention that the total SM background
cross section corresponding to the jet plus missing energy final state
is 3.7~fb, as shown in Table III, and which is 2.42~pb for same final states
as reported in the paper of Ref.~\cite{dreiner}. We observe that our
analysis predicts better sensitivity than others in the 
literature~\cite{tata,pnath}.
Note that in our simulation $\tau$ leptons 
are not considered in final states, which will be simulated in the future
along with a detailed scan of parameter space~\cite{worpro}. 
\\
\section{Acknowledgment}:
The authors wish to thank S. Banerjee and A. Datta for
useful discussion; they also wish to thank G.Majumder and S. Raychaudhuri 
for careful reading of this manuscript. 

\end{document}